\documentclass[conference]{IEEEtran}
\IEEEoverridecommandlockouts
\usepackage{amsmath,amssymb,amsfonts}
\usepackage{silence}
\usepackage{algorithmic}
\usepackage{graphicx}
\usepackage{subfigure}
\usepackage{float}
\usepackage{multirow}
\usepackage{adjustbox}
\usepackage[table]{xcolor}
\usepackage{lipsum}
\usepackage[utf8]{inputenc}
\usepackage[english]{babel}
\usepackage{silence}
\usepackage{csquotes}
\usepackage[
backend=biber,
style=numeric,
sorting=nyt,
]{biblatex}
\newtheorem{finding}{Finding}
\newcommand{\nop}[1]{}
\usepackage{url}
\usepackage{textcomp}
\usepackage{xcolor}
\def\BibTeX{{\rm B\kern-.05em{\sc i\kern-.025em b}\kern-.08em
    T\kern-.1667em\lower.7ex\hbox{E}\kern-.125emX}}

\begin{document}

\nocite{*}

\title{Transaction Characteristics of Bitcoin}
\author{\IEEEauthorblockN{Befekadu G. Gebraselase, Bjarne E. Helvik, Yuming Jiang}
\IEEEauthorblockA{\textit{Department of Information Security and Communication Technology} \\
\textit{NTNU, Norwegian University of Science and Technology, Trondheim, Norway}\\
\{befekadu.gebraselase, bjarne, yuming.jiang\}@ntnu.no}
}

\maketitle
\begin{abstract}
Blockchain has been considered as an important technique to enable secure management of virtual network functions and network slices. To understand such capabilities of a blockchain, e.g. transaction confirmation time, demands a  thorough study on the transaction characteristics of the blockchain. This paper presents a comprehensive study on the transaction characteristics of Bitcoin -- the first blockchain application, focusing on the underlying fundamental processes. A set of results and finding are obtained, which provide new insight into understanding the transaction and traffic characteristics of Bitcoin.
As a highlight, the validity of several hypotheses / assumptions used in the literature is examined with measurement for the first time.  

\end{abstract}


\section{Introduction}
Blockchain has been considered as an important technique to enable secure management of virtual network functions (VNF)~\cite{BVNF} and network slices in 5G and beyond networks~\cite{blockchain5G}. To this aim, understanding the capabilities of the blockchain, e.g. in terms of delay or transaction-confirmation time, is necessary. This naturally demands a thorough study on the transaction characteristics of the blockchain \cite{tranStats2020}, with which, analytical methods (e.g. queueing theory) may be employed to estimate the performance of the blockchain ~\cite{TransactionConfirmationBitcoin}~\cite{ DiscreteBlockchain}.

Surprisingly, even for the first blockchain application, Bitcoin~\cite{Nakamoto}, such a study is still limited. For example, in order to reason the observed exponentially distributed block generation time, two hypotheses on block generation at each miner have been made, namely Bernoulli trial in \cite{BitcoinBlockchainDynamics} and uniform distribution in \cite{trasactionConfirmation}. However, no existing work has investigated whether they both or which one can be justified by the measurement. In addition, among the existing results, e.g. various Bitcoin statistics~\cite{Btc}, block propagation delay~\cite{infoProp2013}, block arrival process \cite{blkArrival2020}, transaction rate and transaction confirmation time~\cite{TransactionConfirmationBitcoin}~\cite{ DiscreteBlockchain}, most are directly generated or derived from the information carried on the Bitcoin blockchain. However, to obtain a deeper understanding of the transaction characteristics of Bitcoin, such information is not enough. For instance, in the literature, Poisson transaction arrival process has been widely assumed, e.g.,~\cite{TransactionConfirmationBitcoin}~\cite{ DiscreteBlockchain}, but due to lack of information on the blockchain about the arrival time of a transaction to a node, the validity of this assumption has never been verified.

The objective of this paper is to report results and findings from an extensive study on the transaction characteristics of Bitcoin, which not only provide answers to the above exemplified open questions but also shed new light on understanding and studying the capabilities of the Bitcoin blockchain. Specifically, the focus is on the most fundamental processes behind Bitcoin, which include the transaction arrival process, the block generation and arrival processes, and the mining pool process. To this aim, a measurement-based study has been conducted, where a dataset has been gathered which contains both information that is globally available from the Bitcoin blockchain, i.e. the ledger, and information that is not available from the ledger but is measured from the local mining pool (mempool). It is worth highlighting that, among these focused processes, the ledger only has timing information for the block generation process, and for the other processes, local measurements are necessary. Based on the collected dataset, an explorative study on the transaction characteristics of Bitcoin has been conducted. 

The results and findings, which constitute the main and novel contributions of this paper, are organized and presented from three angles. {\bf (i)} Firstly, transaction characteristics at the block level, such as block generation, block arrival and block size characteristics, are considered. As a highlight, it is found that, even though the block generation time (at the Bitcoin system level) fits well with an exponential distribution, {\em the two hypotheses on block generation at each miner are both not justified}. Instead, we find another explanation, which is, {\em block generation at major miners has exponentially distributed inter-block generation time}. In addition, distribution-fitting has been conducted on the number of transactions in a block, and block size. The fitting results and findings are also introduced. 
{\bf (ii)} Secondly, transaction level characteristics are focused, which include transaction generation, transaction arrival, transaction size and fee characteristics. Here, {\em the Poisson transaction arrival assumption is examined}. In addition, distribution-fitting has been conducted on transaction size and fee, with results summarized and discussed. 
{\bf (iii)} Thirdly, the dynamics of the mining pool, which underlays the block generation process and relates it to the transaction arrival process, are investigated. In particular, the effect of fee, the fundamental objective of Bitcoin as a digital currency, is included. As a highlight, it is found that {\em the fee-based priority queueing model} assumed in the literature~\cite{TransactionConfirmationBitcoin}~\cite{ DiscreteBlockchain} {\em does not match with the observation}. These results and findings, to the best of our knowledge, has not been previously reported, which provide new insights into understanding the transaction characteristics of Bitcoin.

The rest of this paper is organized as follows. Section~\ref{sec-intro} presents an analysis on and illustrates the workflow of Bitcoin. Then, Section~\ref{sec-dep} introduces the measurement setup and the collected dataset. After that, Section~\ref{sec-inter} introduces results and findings on block level transaction characteristics. Section~\ref{sec-5} presents results and findings on transaction level characteristics. Following that, in Section~\ref{sec-mem}, the dynamics of mempool are focused. In Section~\ref{related}, related work is reviewed. Finally, Section~\ref{sec-con} summarizes the paper.

 \section{Bitcoin Workflow}\label{sec-intro}
Bitcoin is a distributed ledger platform that enables information about transactions to be distributed than centralized, where the ledger is the Bitcoin blockchain that records the transactions. In Bitcoin, all full nodes, also called miners, take part in creating and validating/invalidating transaction blocks and propagating such information, independently~\cite{bitcoinPerf}. Specifically, the users generate transactions for being processed, and the distributed ledger components, i.e. the full nodes or miners, work together to generate and validate transaction blocks and add them to the blockchain.

\begin{figure}[th!]
\centering
  \includegraphics[width=\linewidth,height=0.56\linewidth]{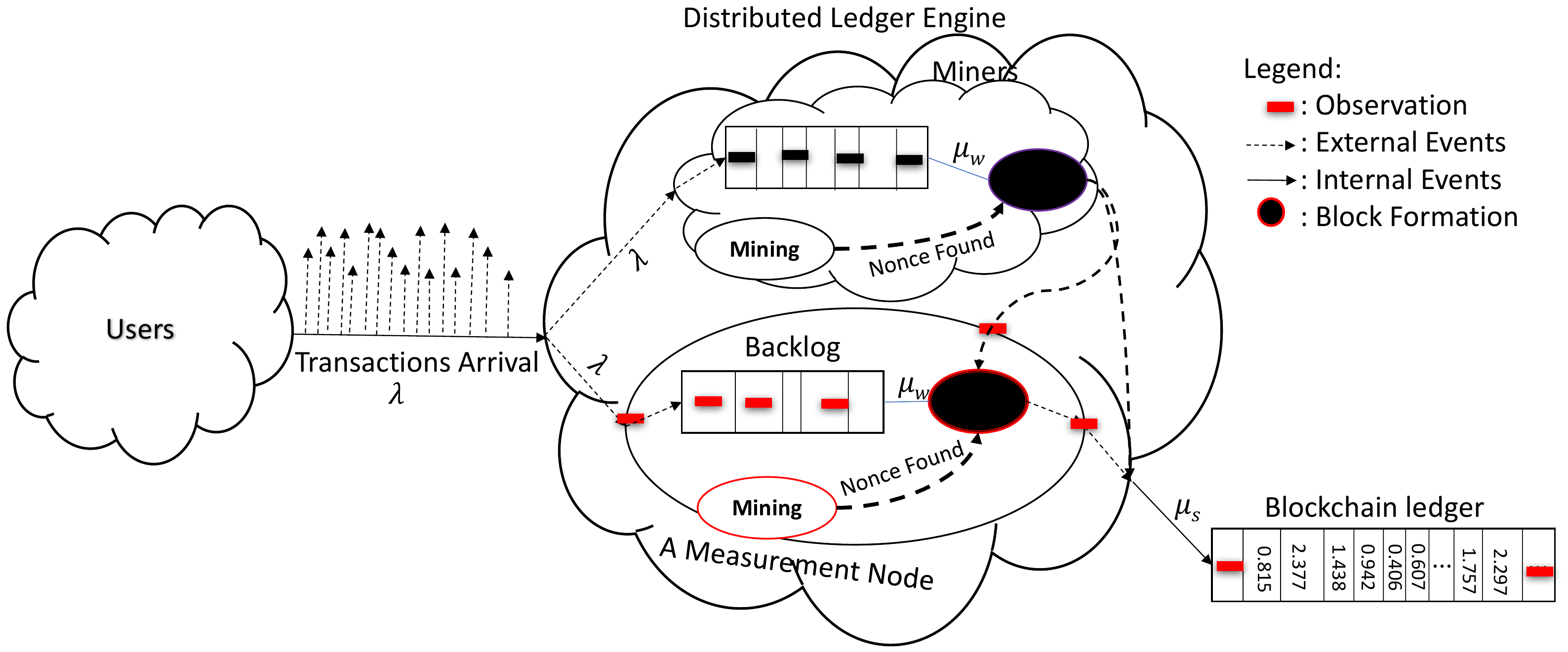}
  \caption{An illustration of the work flow of Bitcoin}
  \label{intro}
\end{figure}

Fig.\ref{intro} illustrates the workflow of transaction arrival, block formation, propagation and validation in Bitcoin. Briefly, after transactions are generated by the users, they are sent to all full nodes for validation. At a full node, upon the arrival of a transaction, the node stores the transaction in its mining pool, called mempool in Bitcoin, waiting for confirmation. In addition, the node may choose unconfirmed transactions in the backlog to pack into a new transaction block, and perform mining to find the mathematical puzzle given by the Bitcoin to gain the right to add the block to the ledger. If the puzzle finding is successful, this newly generated block is added to the blockchain, and this information is sent to all the nodes. At each node, the validity of the newly generated block is checked. If the validity is confirmed with consensus, the updated blockchain is accepted and the transactions in the new block are validated. Such validated transactions are removed from the mempool at each full node that then repeats the above process. Note that, while the above description is brief, the very essence of the workflow, which is needed for the presentation of this paper, is kept. For more details about how Bitcoin works, the original introduction~\cite{Nakamoto} is the best source. 

Accordingly, the workflow involves the following fundamental processes. The first is the transaction arrival processes, i.e. the process of transaction arrivals to a full node. The next is the block generation process, i.e. the process of block generations in the Bitcoin network. The third is the block arrival process, i.e. the process of block arrivals to a full node. The fourth is the mining pool process that records the backlog dynamics of the mempool at a full node.  

In order to provide a comprehensive study on the transaction characteristics of Bitcoin, all these processes should be considered. However, it is easily checked that the information carried on the ledger is not enough. To address this limitation, additional information has been collected through measurement. The setup and the collected dataset are described in the next section.

\section{Measurement Setup and Dataset Summary} \label{sec-dep}  

\begin{figure}[th!] 
\centering
  \includegraphics[width=\linewidth,height=0.75\linewidth]{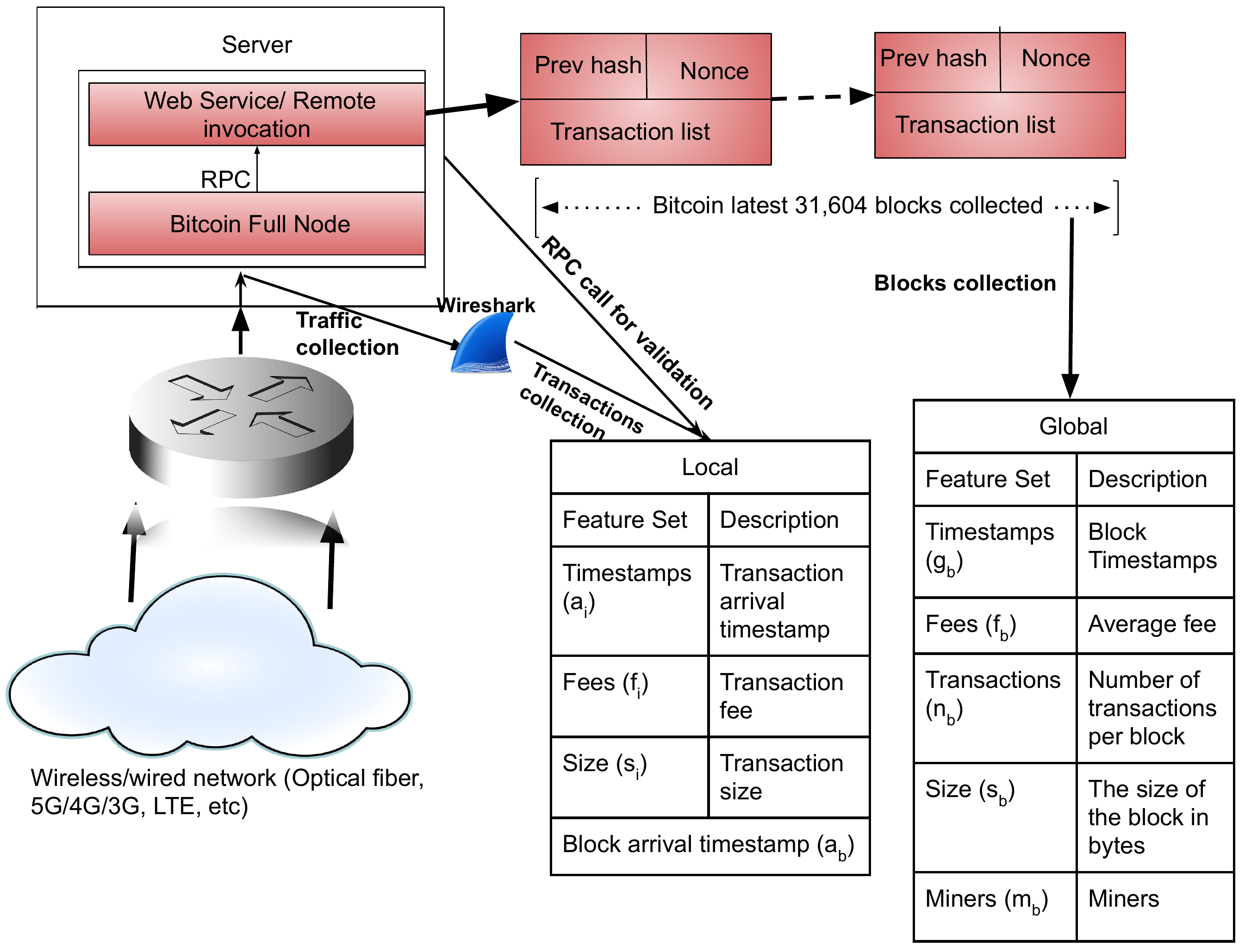}
  \caption{Testbed deployment and dataset attributes}
  \label{deployment}
\end{figure}

For the measurement study, a testbed as shown in Fig.~\ref{deployment}, has been implemented to record information about Bitcoin transactions. The testbed includes a server installation of a full Bitcoin node, and this full node is essentially a measurement node as illustrated in Fig.~\ref{intro}. 

Through the testbed, a dataset, consisting of two parts, has been collected. One part of the dataset records information from the ledger that is globally available, called the {\em global information part}. Another part records locally available information about each transaction and block as well as the backlog status of the mempool. This part is called the {\em local information part}. The measurement period of the dataset is from 7th March 2019 till 3rd October 2019, and the dataset consists of over 79 million transactions contained in 31 thousand blocks on the ledger and recorded at the installed full node. 

The ledger dataset was collected through a REST API that enables RPC calls to the installed node to collect information about blocks and transactions. The mempool dataset was collected through Wireshark that collects traffic information from the network interface of the node, while RPC calls to the installed node were done to validate that the extracted transaction is available at the mempool.  To do so, we used a C++ code to act as a middleman between the installed node and traffic collection from the interface, as demonstrated in Fig.~\ref{deployment}.  

The recorded information in the global information part of the dataset includes, for each block $b$ on the blockchain, the number of transactions ($n_b$) in the block, the block generation time ($g_b$), its miner ($m_b$), the size of the block ($s_b$), and the fee ($f_b$). The locally recorded information from the installed full node includes for each transaction $i$, the arrival time timestamps ($a_i$), the transaction fee ($f_i$), and the size ($s_i$), and additionally for each block $b$, its arrival time ($a_b$). A brief summary of these focused features is also shown in Fig.~\ref{deployment}.

In the literature, several platforms provide similar datasets. However,  the data extracted from such a source lacks some information that is available in ours. For instance, the set of mempool features, timestamp ($a_i$), transaction fee ($f_i$), and size ($s_i$), which are related to transaction arrivals, are unique in our dataset which generally is not available form the literature platforms. With such information, we can extract the number of bytes that arrive at the mempool in an interval. Additionally, some more detailed information related to each block, which is gathered from the installed full node in our testbed, is not available in the other sources. In particular, in each block, there are many transactions, and each transaction has a number of attributes such as size, fee, and timestamp. Such detailed information cannot be found from outside sources: What is available there is only some piece of general information. Table~\ref{sourcecomparison} provides a comparison of what transaction and block attributes are included in the several well-known platforms and ours, where {\em IIK testbed} represents our testbed.

\begin{table}[ht!]
\caption{Data source comparison} 
\centering 
\begin{adjustbox}{width=0.48\textwidth}

  \begin{tabular}{|l|l|l|l|l|l|l|l|l|l|l|}
    \hline
    \multirow{2}{*}{Dataset} &
      \multicolumn{4}{c|}{Locally recorded attributes} & 
      \multicolumn{4}{c|}{Block attributes} \\
     & $a_i$ & $f_i$ & $s_i$& $a_b$ & $g_b$ & $f_b$ & $n_b$ & $s_b$\\
    \hline
    Blockstream \cite{Blockstream}  & $\times$ &  \checkmark & \checkmark  &  $\times$ & \checkmark & $\times$ & \checkmark& \checkmark \\
    \hline
    Bitaps  \cite{Bitaps}  & $\times$ &  $\times$ & $\times$ & $\times$ & $\times$ & \checkmark & \checkmark& \checkmark \\
    \hline
    Btc \cite{Btc}  & $\times$ &  $\times$ & $\times$  &  $\times$ &\checkmark & $\times$ & \checkmark& \checkmark \\
    \hline
    Explorers \cite{Explorer} &$\times$ & $\times$ & \checkmark &  $\times$& $\times$ & \checkmark & \checkmark& \checkmark \\
    \hline
    IIK testbed & \checkmark & \checkmark & \checkmark  & \checkmark &\checkmark & \checkmark & \checkmark & \checkmark \\
    \hline
  \end{tabular}
  \end{adjustbox}
 \label{sourcecomparison}
\end{table}

\section{Block-Level Characteristics } \label{sec-inter} 

In this section, a number of transaction characteristics at the block level are investigated, which are related to block generation and arrival time processes, the number of transactions in a block, and block size.

\subsection{Inter-Block Generation Time} 
We first focus on the block generation process and examine the two hypotheses in the literature \cite{BitcoinBlockchainDynamics}  and \cite{trasactionConfirmation} in trying to reason the observed inter-block generation time distribution. 

The Bitcoin system uses the UTC +1 zone to synchronize full nodes. Using the same timezone among nodes helps to reduce wrong interpretation or modification of information to a different order.  At the generation of a block $b$, its generation time $g_b$ is added to the block. In this way, the Bitcoin blockchain keeps track of block generations in the system.

\begin{figure}[ht!]
\centering
  \includegraphics[width=\linewidth, height=0.45\linewidth]{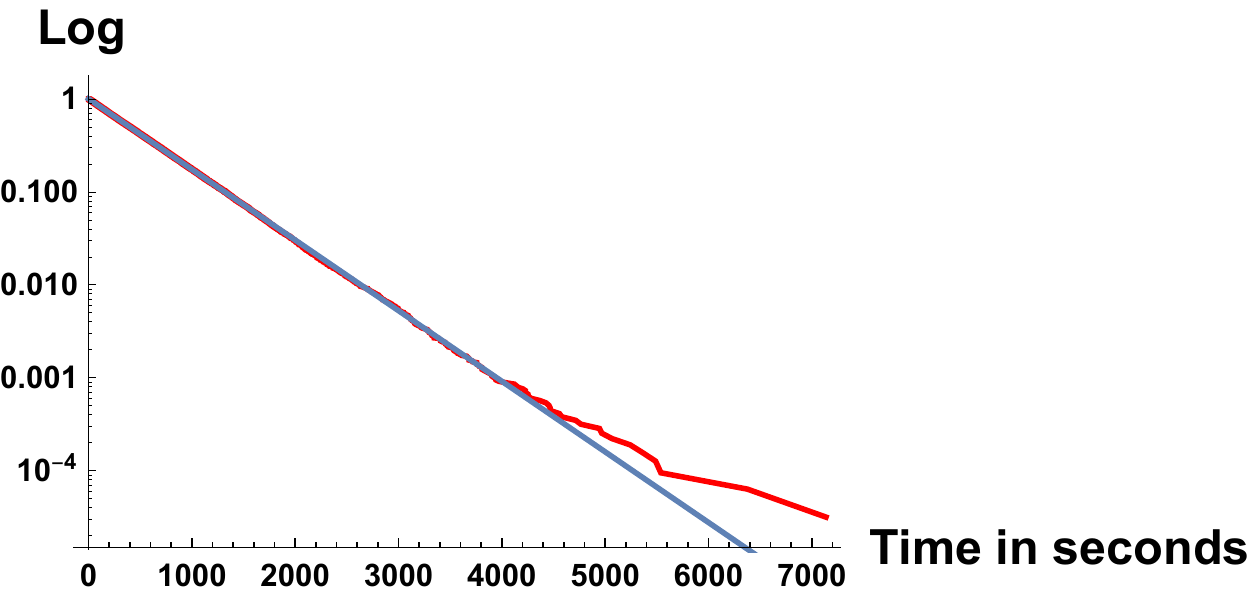}
   \caption{The inter-block generation time and the inter-block arrival time, fit to an exponential distribution}
  \label{block_gen}
\end{figure}

Fig.~\ref{block_gen} shows that the inter-block generation time of Bitcoin can be excellently matched with a negative exponential distribution, as also reported in the literature \cite{BitcoinBlockchainDynamics}\cite{trasactionConfirmation}, event though there is some deviation at the tail likely attributed to the total number of blocks in the measurement period.  

\begin{figure}[th!]
    \centering
    \subfigure[Mining pool contribution in terms of generating blocks]
    {
    \includegraphics[width=0.45\linewidth, height=0.3\linewidth]{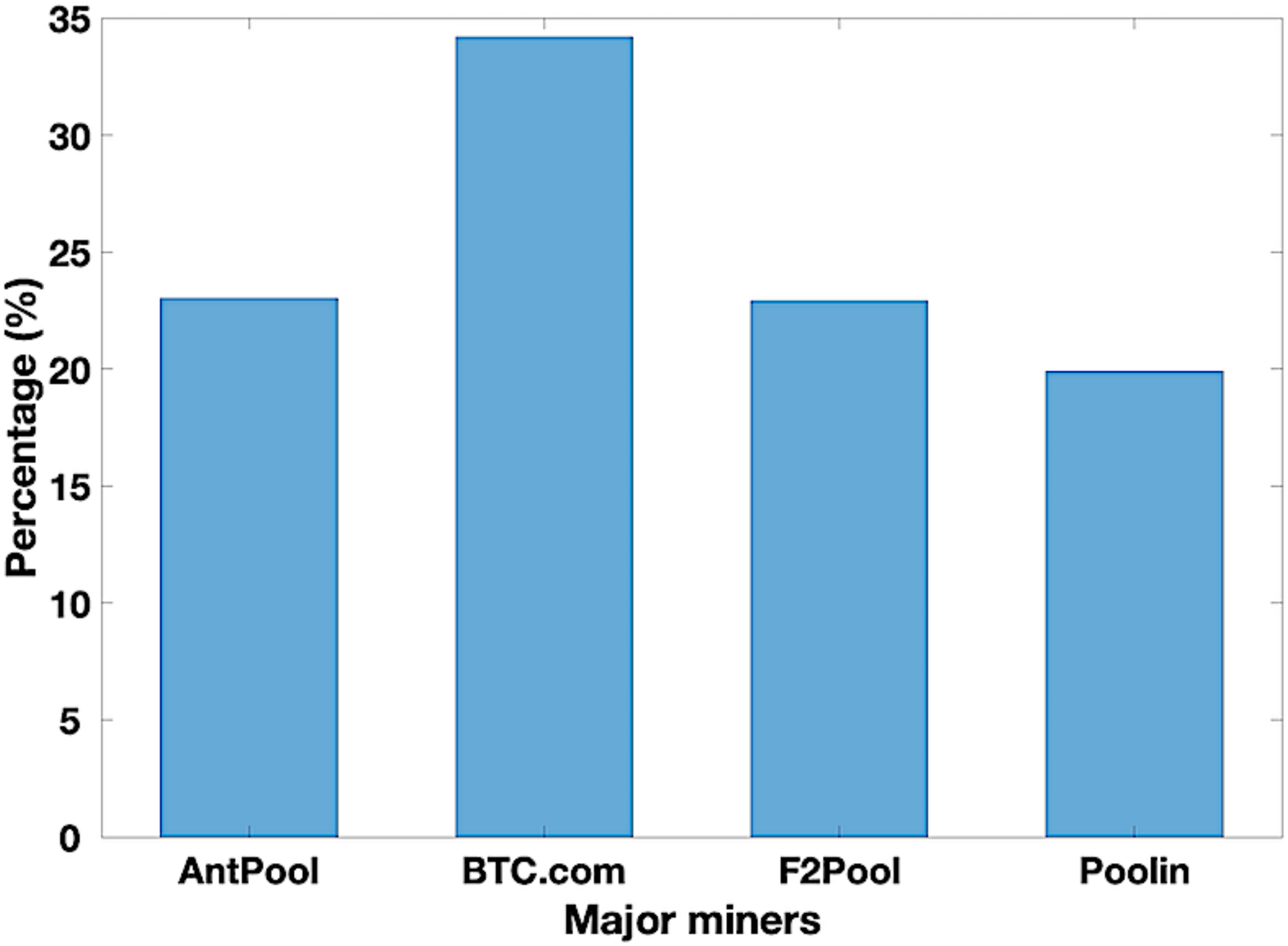}
   \label{miners}
    }
    \subfigure[Inter-block arrival time distribution at a miner]
    {
    \includegraphics[width=0.45\linewidth,height=0.3\linewidth]{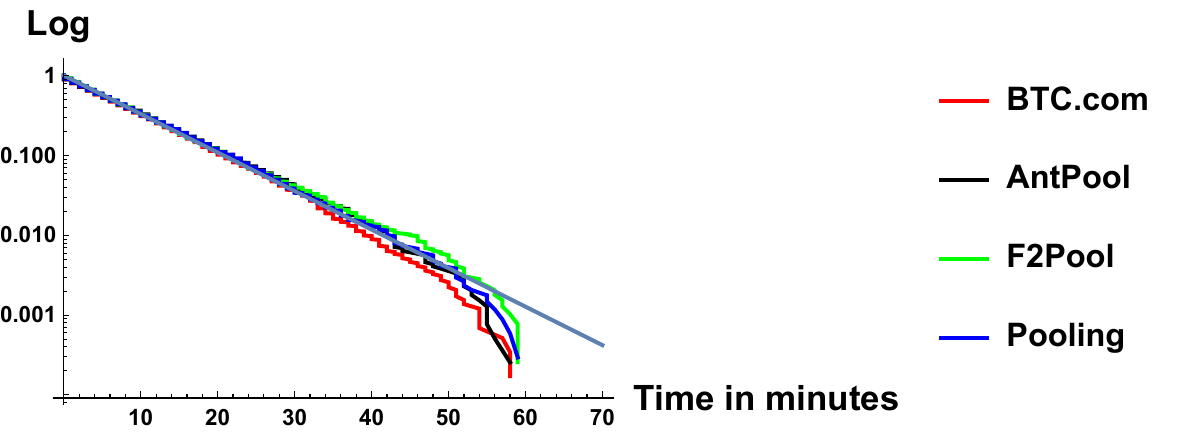}
   \label{Antminersinter}
    }
    \label{kstest}
  \caption{Block contribution by miners and per-miner inter-block generation}  
\end{figure}

To further find explanation for the exponentially distributed inter-block generation time, we investigate this distribution of each miner. To this aim, the contributions of main miners to block generation is first examined and the results are shown in Fig.~\ref{miners}. The figure shows that the majority (80\%) are contributed by the few top private miners including Antipool, BTC, BTC.Top, BitFury, F2pool, and viaBtc. In addition, some public mining pools such as Pooling exist, used by nodes to participate in the mining pool. We also observed that BTC.com, AntPool, F2Pool, and Pooling mining pools contributes the majority of the blocks to the ledger. The number of blocks generated by these pools are not evenly distributed, while the few major miners take most of the valid block.

Based on Fig.~\ref{miners}, several major contributing miners are chosen and Fig.~\ref{Antminersinter} reports their inter-block generation time distribution. Also in Fig.~\ref{Antminersinter}, an exponential distribution fitting curve, the straight line, is presented. As can be observed from the figure, the inter-block generation time can be well approximated by an exponential distribution. This is in contradiction with the two hypotheses found in \cite{trasactionConfirmation} and \cite{BitcoinBlockchainDynamics}.

Note that, it is well known that the aggregation of independent Poisson processes also results in a Poisson process, or in other words, the aggregate point process of independent point processes, each of which has exponentially distributed inter-arrival time, also has exponentially distributed inter-arrival time. It is then worth highlight that, the finding in Fig.~\ref{Antminersinter} provides a previously unreported explanation for the exponentially distributed inter-block generation time in the Bitcoin system, i.e. it is resulted from similar distributions at the miners.

\begin{finding}
{\bf The exponentially distributed inter-block generation time on the blockchain is likely attributed to the exponentially distributed inter-block generation time at major miners. }
\end{finding}

\subsection{Inter-Block Arrival Time}

It is worth highlighting that the block arrival process to a node is different from the block generation process of the Bitcoin system. This is due to that after the generation of a new block, the updated ledger containing the new block needs to be propagated through the Bitcoin network to each node. This causes propagation delay from the generation of each block at its miner to the arrival of the block to a node, $a_b - g_b$. 

\begin{figure}[ht!]
    \centering
    \subfigure[The block inter-arrival time fitting to a n.e.d]
    {
    \includegraphics[width=0.45\linewidth, height=0.3\linewidth]{interarrivaltime.pdf}
   \label{block_arrival}
    }
    \subfigure[Block propagation delay distribution fitting to a n.e.d]
    {
    \includegraphics[width=0.45\linewidth,height=0.3\linewidth]{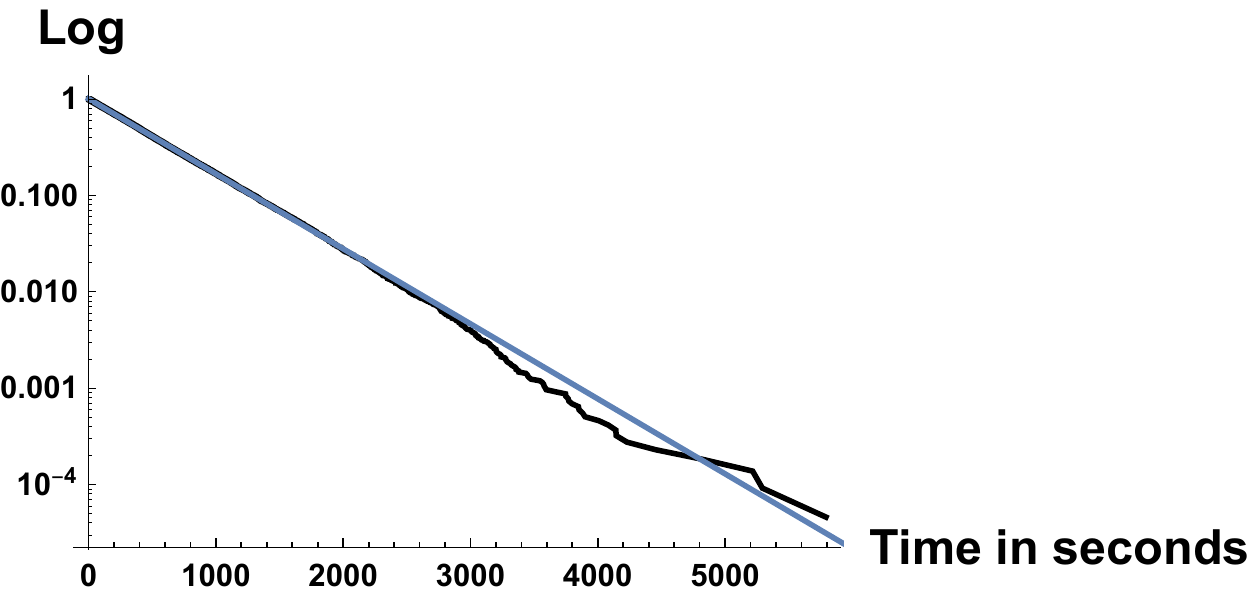}
   \label{block_prop}
    }
    \label{block-a-p}
  \caption{Block arrival time and block propagation delay}  
\end{figure}

In the literature, e.g. \cite{infoProp2013}, it has been discussed and conjectured that the block propagation delay is exponentially distributed, but the conjecture is not examined with measurement. We have also performed analysis on the propagation delay with our collected measurement dataset. Based on the arrival time $a_b$ recorded at our node and its generation time $g_b$, we have found an average of 53 seconds for the block propagation delay. Its distribution is shown in Fig.~\ref{block_prop}. It can be observed from the figure that the block propagation delay well fits an exponential distribution, validating the conjecture in \cite{infoProp2013}.

\begin{figure}[t!]
    \centering
    \subfigure[K-S test for block inter-arrival time distribution where D represents the maximum distance between the exponential and empirical CDF]
    {
    \includegraphics[width=0.45\linewidth, height=0.4\linewidth]{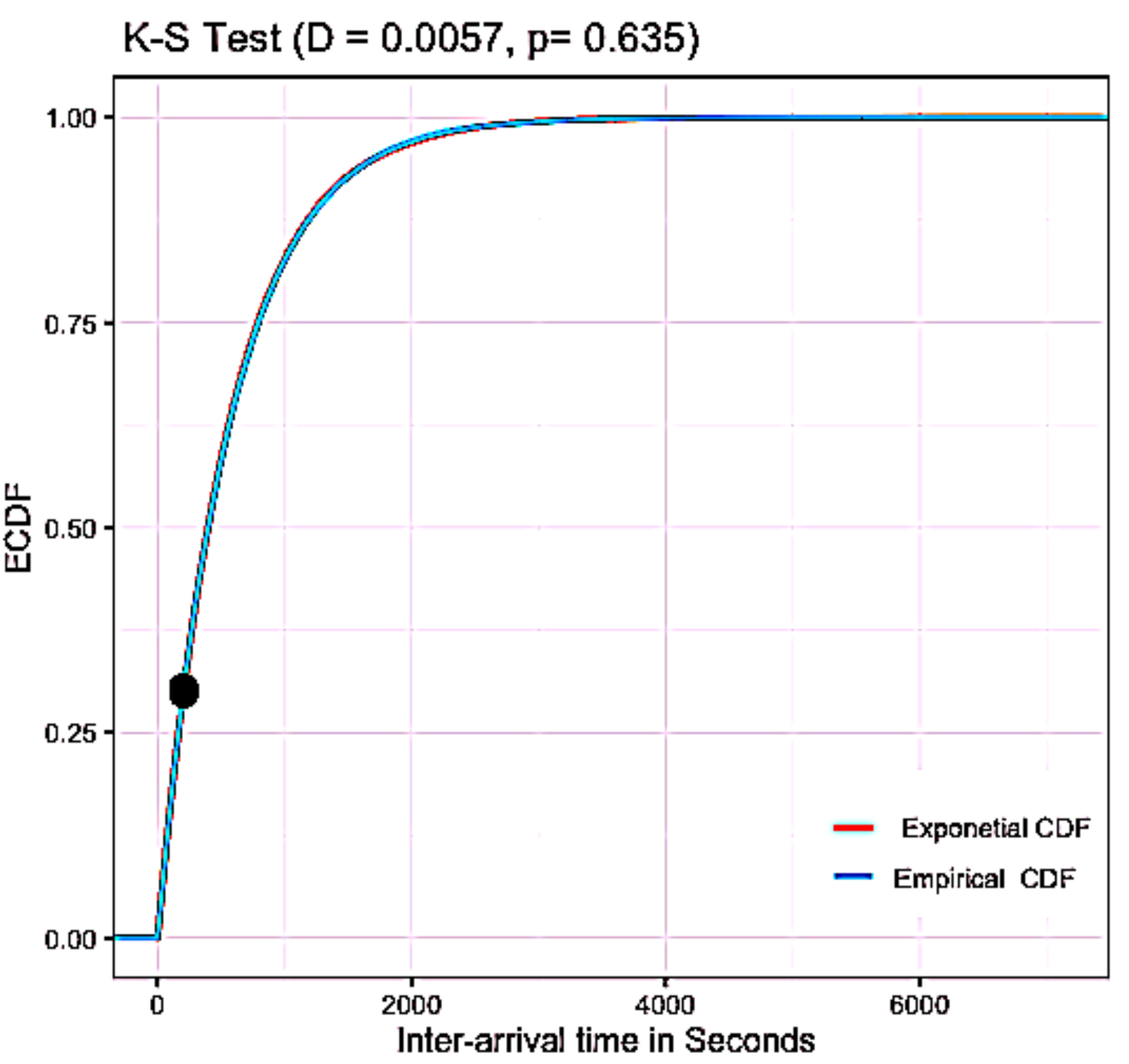}
   \label{testCase}
    }
    \subfigure[The autocorrelation of block inter-arrival time in seconds]
    {
    \includegraphics[width=0.45\linewidth,height=0.4\linewidth]{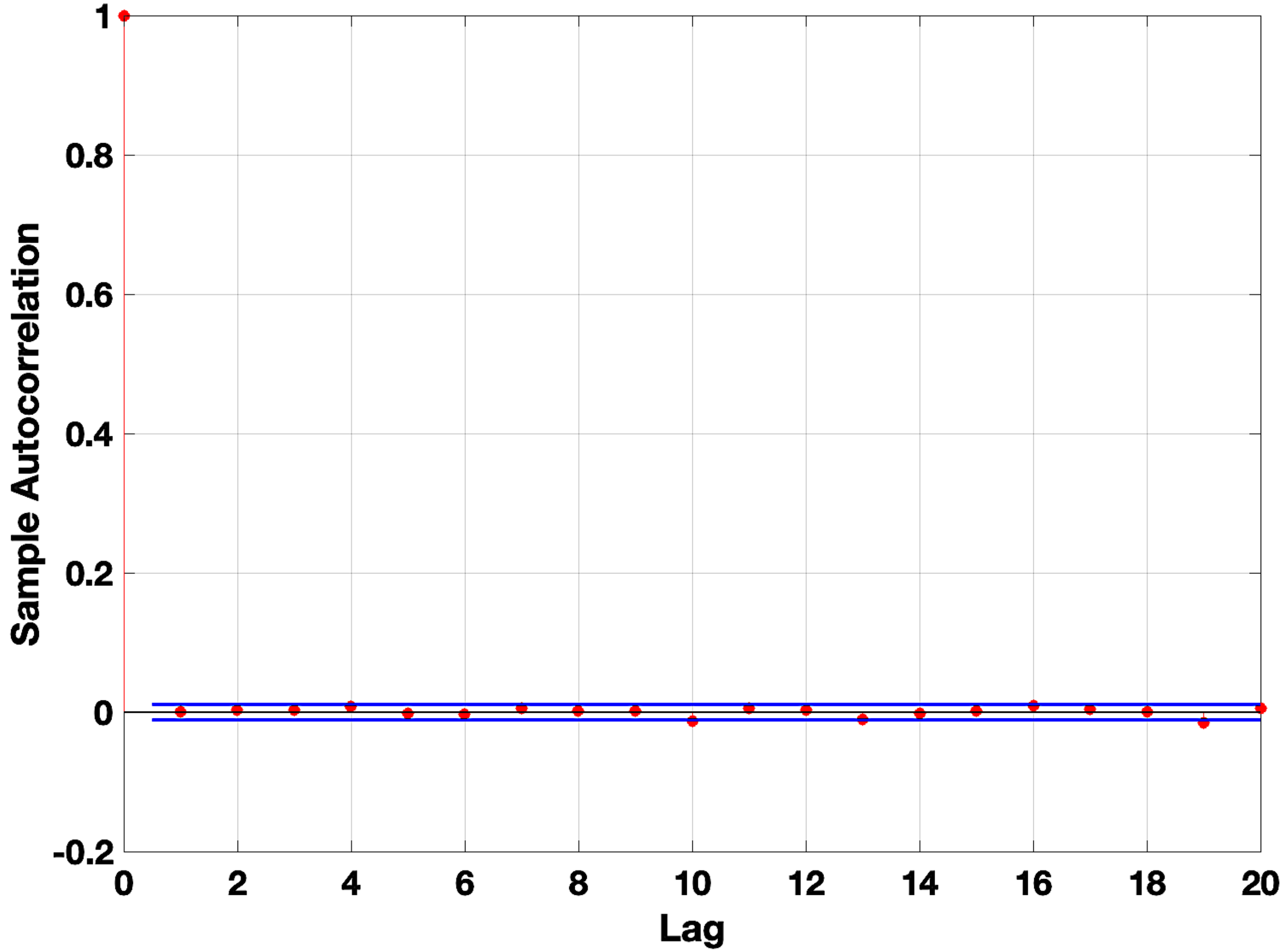}
   \label{auto} 
    }

  \caption{K-S test and autocorrelation}  
\end{figure}

\nop{
\begin{figure}[ht!] 
\centering
 \includegraphics[width=0.5\linewidth]{Rplot01.pdf}
   \caption{K-S test for block inter-arrival time distribution where D represents the maximum distance between the exponential and empirical CDFs (\textcolor{red}{In the figure, remove "p". })}
  \label{testCase}
\end{figure} 
}

For the inter-block arrival time between two adjacent blocks $b_i$ and $b_{i+1}$, it can be calculated from their arrival times recorded in the local information, i.e. $a_{b_{i+1}}-a_{b_i}$. Its distribution is shown in Fig.~\ref{block_arrival}. As can be observed from Fig.~\ref{block_arrival}, the distribution can be well approximated by an exponential distribution. This appealing finding can indeed be expected from the distribution of inter-block generation time and the distribution of propagation delay due to the following relationship between them: 
\begin{equation*}
(a_{b_{i+1}}-a_{b_i}) = (g_{b_{i+1}}-g_{b_i}) +[(a_{b_{i+1}} - g_{b_{i+1}}) - (a_{b_{i}} - g_{b_{i}})]
\end{equation*}
where, on the right side, the inter-block generation time $g_{b_{i+1}}-g_{b_i}$ is approximately exponentially distributed as discussed in the previous subsection, and the second term is the propagation delay difference. Since propagation delay is also approximately exponentially distributed, the difference shown as the second term can be approximated to have a Laplace distribution, from the well-known result of difference of two exponentially distributed random variables. Furthermore, from the sum of exponential and Laplace distributions \cite{Cahillane2020}, an exponential decay in the inter-block arrival time is expected. 

In addition to a K-S test confirming the excellent match, which is shown in Fig.~\ref{testCase}, we have also examined if blocks arrive independently. This is done by checking the autocorrelation of the block arrival time series, under different time lags. A summary of the autocorrelation values is presented in Fig.~\ref{auto}. As can be seen from the table, the autocorrelation is close to zero under all these lags with the largest difference only around $1\%$, which is an indication that block arrivals are not correlated. 

\begin{finding}
{\bf The block arrival process to a node approximately has an exponentially distributed inter-block arrival time with independent block arrivals. }
\end{finding}

\subsection{Number of Transactions in a Block and Block Size} \label{sec-dat} 

In  contrast to very few results about inter-block generation / arrival time distributions, the literature has a lot of results about $n_b$ -- the number of transactions in a block and $s_b$ -- the size of a block, such as those reported in the various platforms \cite{Blockstream}  \cite{Bitaps} \cite{Btc} \cite{Explorer}.  In this and the subsequent subsections, we report results that are either with more detailed information or from new different perspectives.

\subsubsection{Distributions of $n_b$ and $s_b$} \label{fit}
Distribution fitting has been conducted to investigate what distributions fit $n_b$ and $s_b$. Specifically, maximum likelihood estimation (MLE) based on the Kolmogorov–Smirnov (K-S) goodness-of-fit test and the Anderson–Darling (A-D) test has been applied. The K-S test value represents the maximum distance between the empirical and the tested distribution's cumulative distribution functions (CDFs), and the A-D value measures the area between the fitted line based on the tested distribution and the empirical distribution functions. Typically, a smaller K-S value indicates a better fitted distribution, which is similarly for the A-D test. 

The Fitdistrplus~\cite{fitdistrplus} and fitdist~\cite{Matlab} packages from R and Matlab are utilized to conduct the distribution fitting and obtain the corresponding parameter values of the distribution.  For $n_b$ and $s_b$, it is found that logistic, t location-scale, and normal distributions fit better than other distributions. Table~\ref{closefit_arrival} lists the distributions and the corresponding parameters' values, where the $\mu$ and $\sigma$ respectively represent the mean and standard deviation, and the extra $v$ for t location-scale distribution is its degree of freedom parameter. We randomly selected twenty thousand blocks and the number of transactions inside to find a close fit.
As can be observed from the table, the number of transactions in a block and the block size both have the logistic distribution as the best fit. However, between the two, the fitting of $n_b$ is much better. In particular, it seems $n_b$ can be reasonably approximated to have a logistic distribution.

\begin{table}[ht!]
\caption{Distribution fit of $n_b$ and $s_b$} 
\centering 
\begin{adjustbox}{width=0.48\textwidth}

  \begin{tabular}{|l|l|l|l|l|l|l|l|l|}
    \hline
    \multirow{2}{*}{Distribution} &
      \multicolumn{3}{c|}{Number of transactions per block ($n_b$)} &
      \multicolumn{3}{c|}{Block size ($s_b$)} \\
     & $KS$ & $AD$ & $Parameter$ & $KS$ & $AD$ & $Parameter$ \\
    \hline
    Logistics  & 0.0079 &  670.8 & $\mu$= 1.18, $\sigma$= 0.23  & 0.13 &  1898.6 & $\mu$= 1.168, $\sigma$= 0.16  \\
    \hline
    Normal  & 0.12 &  891.6 & $\mu$= 1.13, $\sigma$= 0.41  & 0.24 & 2764.9 &$\mu$= 1.2075, $\sigma$= 0.3395 \\
    \hline
    t-location  & 0.27  &  4119.7 & $\mu$= 1.12, $\sigma$= 0.35, v=7.8  & 0.3 & 6171.9 & $\mu$= 1.22, $\sigma$= 0.096, v=1.67 \\
    \hline
   
  \end{tabular}
  \end{adjustbox}
 \label{closefit}
\end{table}

\subsubsection{Correlation between $n_b$ and $s_b$}  

\begin{figure}[ht!]
    \centering
    \subfigure[The joint PDF of $n_b$ and $s_b$]
    {
    \includegraphics[width=0.45\linewidth, height=0.5\linewidth]{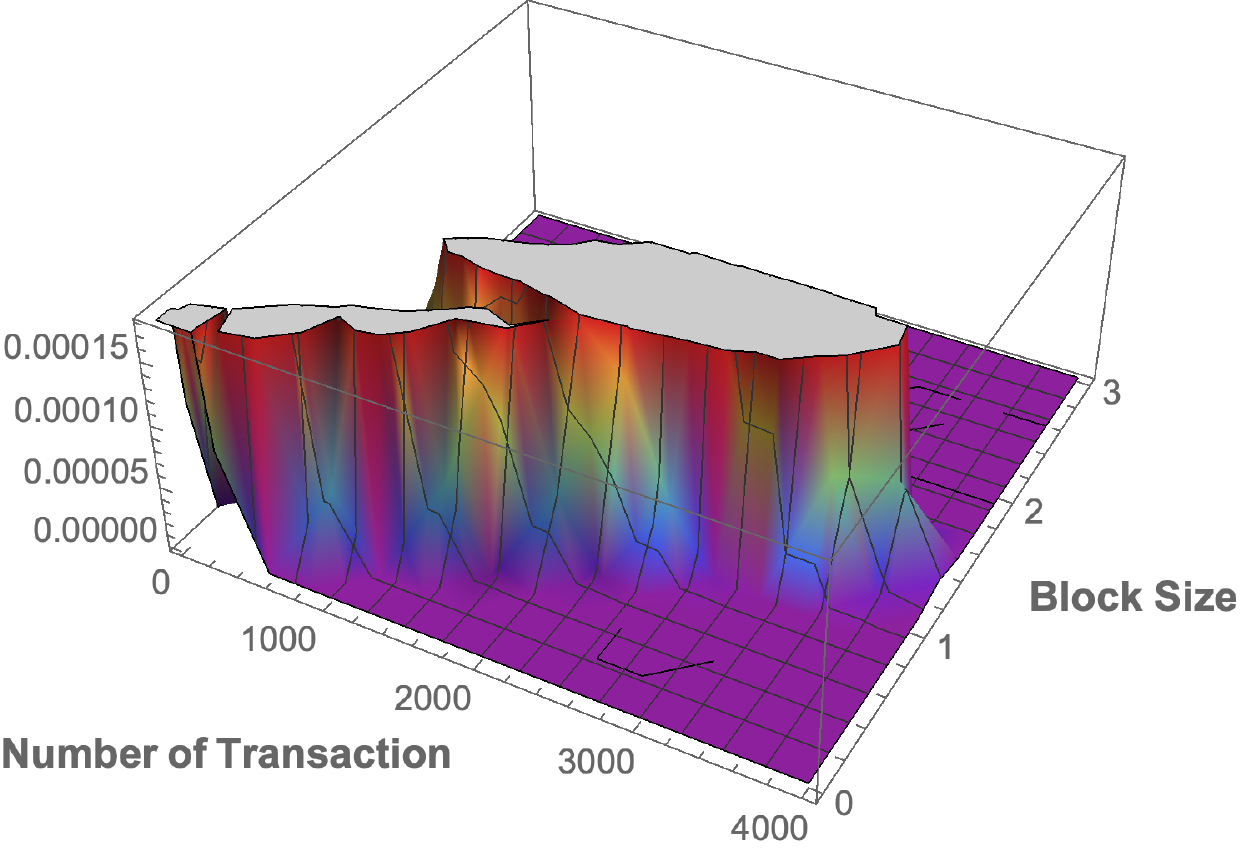}
   \label{numSize}
    }
    \subfigure[Pearson's correlation coefficient]
    {
    \includegraphics[width=0.45\linewidth,height=0.5\linewidth]{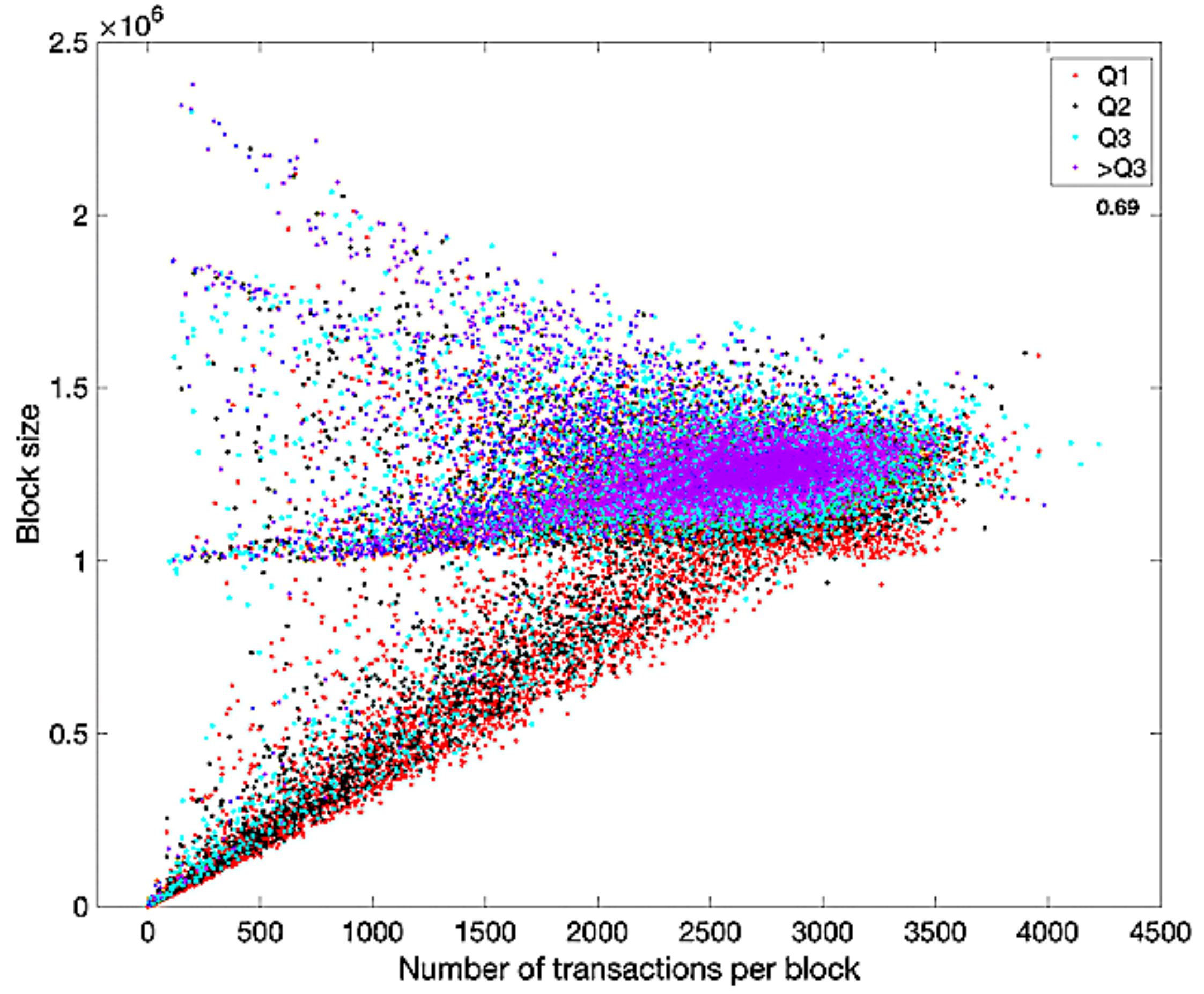}
   \label{fig:corr}
    }
    \label{n-b-cor}
  \caption{Correlation between $n_b$ and $s_b$}  
\end{figure}

Fig.~\ref{numSize} illustrates the joint PDF of $n_b$ and $s_b$. As we can see from the figure, the dependence between the two variables varies. In general, a larger block has a higher number of transactions included. While this is as expected, Fig.~\ref{numSize} details this relationship.   
In addition, Fig.~\ref{fig:corr} illustrates the scatter diagram of Pearson's correlation coefficient between the size of a block, $s_b$, and the number of transactions in the block,  $n_b$. It also demonstrates the relationship of $s_b$ and $n_b$ for Q1 (25\%), Q2 (50\%), Q3 (75\%), and greater than Q3 ($>$Q3) for $f_b$. These intervals are (0,Q1), (Q1,Q2), (Q2,Q3), and (Q3,$\infty$).
As can be observed from Fig.~\ref{fig:corr}, there is strong correlation between $s_b$ and $n_b$: There is a clear pattern shown by the correlation scatter points.

Specifically, it is visible that having a large $n_b$ often implies a higher chance of being in a bigger $s_b$, as illustrated by blue and bold black dotted blocks when the $n_b$ higher than 2000, even though some high size blocks have a small number of transactions in the block.  Sometimes, the number of transactions waiting for confirmations is smaller than the block size capacity; in such cases, we will see blocks filled with fewer numbers than the expected.  In  Fig.~\ref{fig:corr},  we can see the black and red dotted straight line around 0 - 1 MB, indicating generating a block not filled with a maximum capacity as the consequence of the mempool containing a small number of transactions waiting. On the other hand, we can also observe a horizontal line around the $s_b$ 1 - 1.5 MB and where $n_b$ is more significant than 2000, which indicates more transactions waiting while the block filled to the maximum limit.  Additionally, we can also see pink and light-green colored blocks with a small number of transactions in a block while the size is pushed to the maximum limit.  

Furthermore, we can also observe that the average gain of miners playing a crucial role. The blocks with a higher average fee per block ($>$Q3) contain a higher gain; on the other hand, most less-filled black and red colored blocks contain less average gain.

\begin{finding}\label{f-3}
{\bf There is positive, strong, and nonlinear correlation between the size of a block and its number of transactions. }
\end{finding}

As a remark, the Bitcoin community has pushed to increase the block size $s_b$ to 1 MB. The soft-fork extension of the $s_b$ is to add more transactions to each block.  Since there is no policy to force the miner from adding transactions of the interest, we can see from Fig.~\ref{fig:corr} that a large block can be filled with a small number of transactions, $n_b$. In most cases, when the block is filled with small $n_b$, it exhibits that the size is pushed to the maximum limit from 2 - 2.5 MB.  At the same time, we can also observe that block size is highly concentrated around 1 - 1.5 MB, in which case and maybe because legacy nodes are still operational.

\subsection{Characteristics in Different Time Periods} 

\begin{figure}[ht!]
    \centering
    \subfigure[$n_b$ characteristics]
    {
    \includegraphics[width=0.45\linewidth, height=0.5\linewidth]{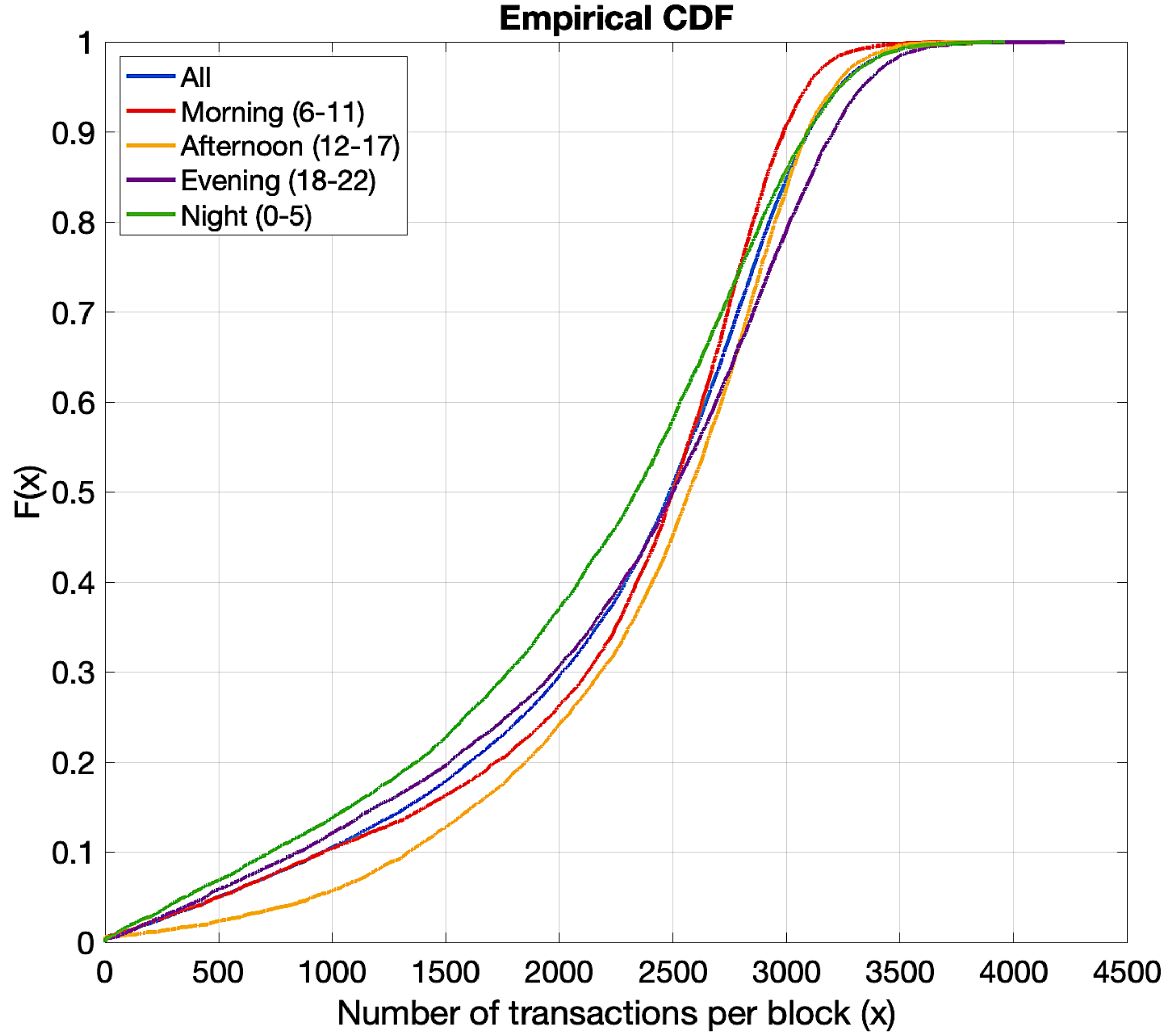}
   \label{dayandnight}
    }
    \subfigure[$s_b$ characteristics] 
    {
    \includegraphics[width=0.45\linewidth, height=0.5\linewidth]{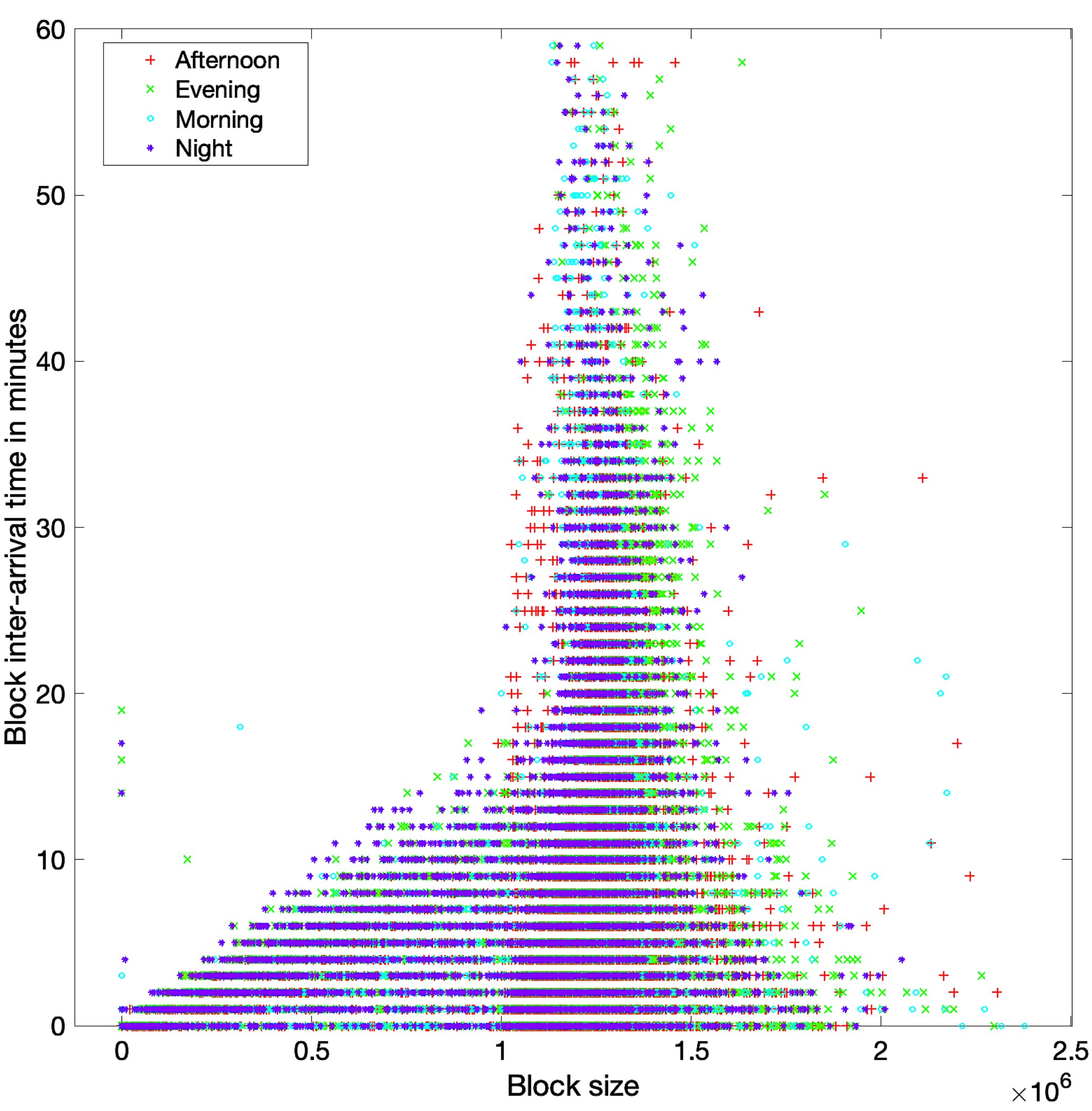}
   \label{blockbyday}
    }
    \label{time-eff}
  \caption{$n_b$ and $s_b$ characteristics in different time periods}  
\end{figure}

We are interested in finding if and how $n_b$ and $s_b$ may differ in different time periods. As the CDF of $n_b$ reported in Fig.~\ref{dayandnight},  in the morning and evening, a block holds on average 2500 transactions, and in the night and afternoon, a block contains no more than 3300 transactions in 90\% of cases. Still, in all the cases, it can grow larger than 3500 in 1\% of the cases.

Fig.~\ref{blockbyday} reports that $s_b$'s distribution also varies as $n_b$.  In the afternoon and evening, the $s_b$'s ranges are higher than in the morning and night.  The $s_b$ in the evening is relatively larger than in other periods. This may be due to a higher $n_b$ in the evening.  In the morning and evening, the number of blocks are generated less frequently, i.e. with higher inter-block generation time shown in the figure, than the rest of the day.

\begin{figure}[t!]
    \centering
    \subfigure[Working days]
    {
    \includegraphics[width=0.8\linewidth,height=0.52\linewidth]{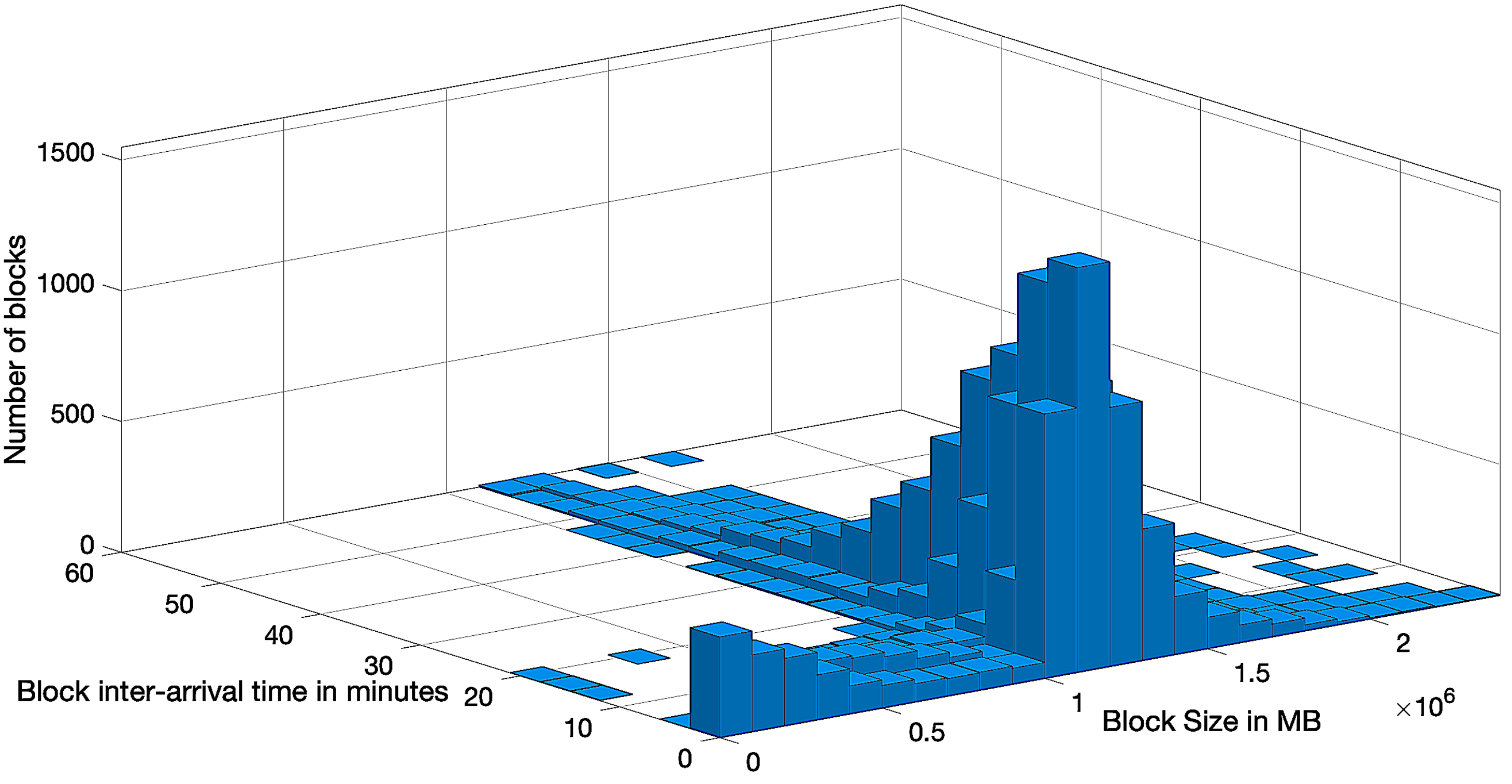}
   \label{block_arrival_size}
    }
    \subfigure[Weekend]
    {
    \includegraphics[width=0.8\linewidth,height=0.52\linewidth]{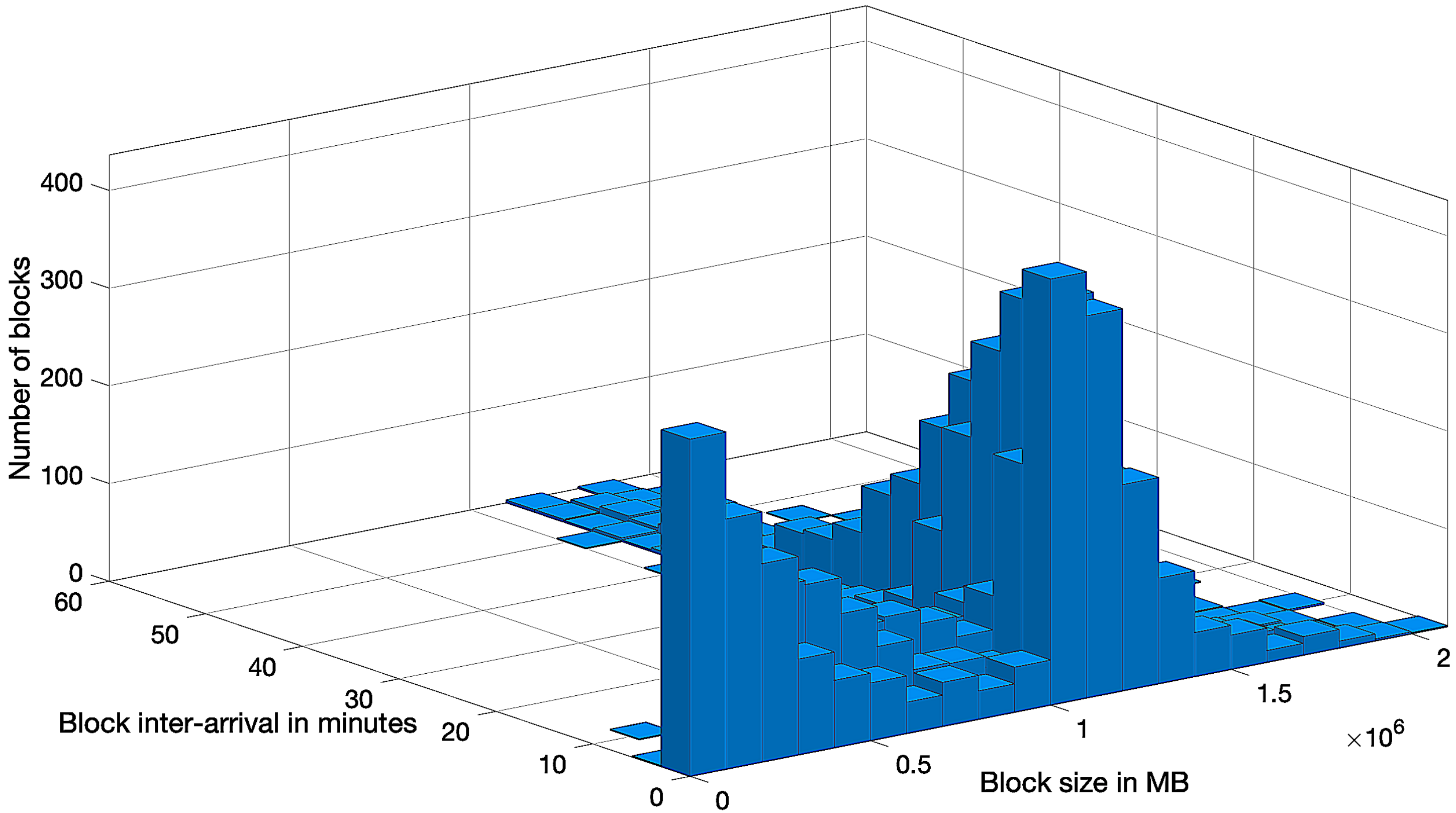}
   \label{block_arrival_weekend}
    }
    \label{kstest_arrival}
  \caption{Block generation in working and weekend days} 
\end{figure}

Fig.~\ref{block_arrival_size} and Fig.~\ref{block_arrival_weekend} further show how $s_n$'s  distribution varies over the working and weekend days.  In the working days, the $s_n$  is more concentrated over the range of 1 to 1.5 MB, and there are 9229 blocks arrival with an inter-generation time of less than 5 minutes. However, in the weekend days, $s_n$ stands between 0.2 to 1.8 MB, and about 3700 blocks are found with an inter-generation time of less than 5 minutes.

\begin{finding}\label{f-5}
{\bf The characteristics of block size and number of transactions can differ significantly in different time periods. }
\end{finding} 

\section{Transaction-Level Characteristics} \label{sec-5}
In Bitcoin's design, a transaction confirmation time of 10 minutes is inherent \cite{Nakamoto}. Based on the arrival time $a_b$ recorded at our node and its generation time $g_b$, we have found that on average a transaction needs 600 seconds ($T_w$) from it is received by the Bitcoin system till the corresponding block is generated, i.e. the transaction is confirmed then. This confirms the design principle of Bitcoin. 

\begin{figure}[t!]
\centering
     \includegraphics[width=\linewidth,height=0.7\linewidth]{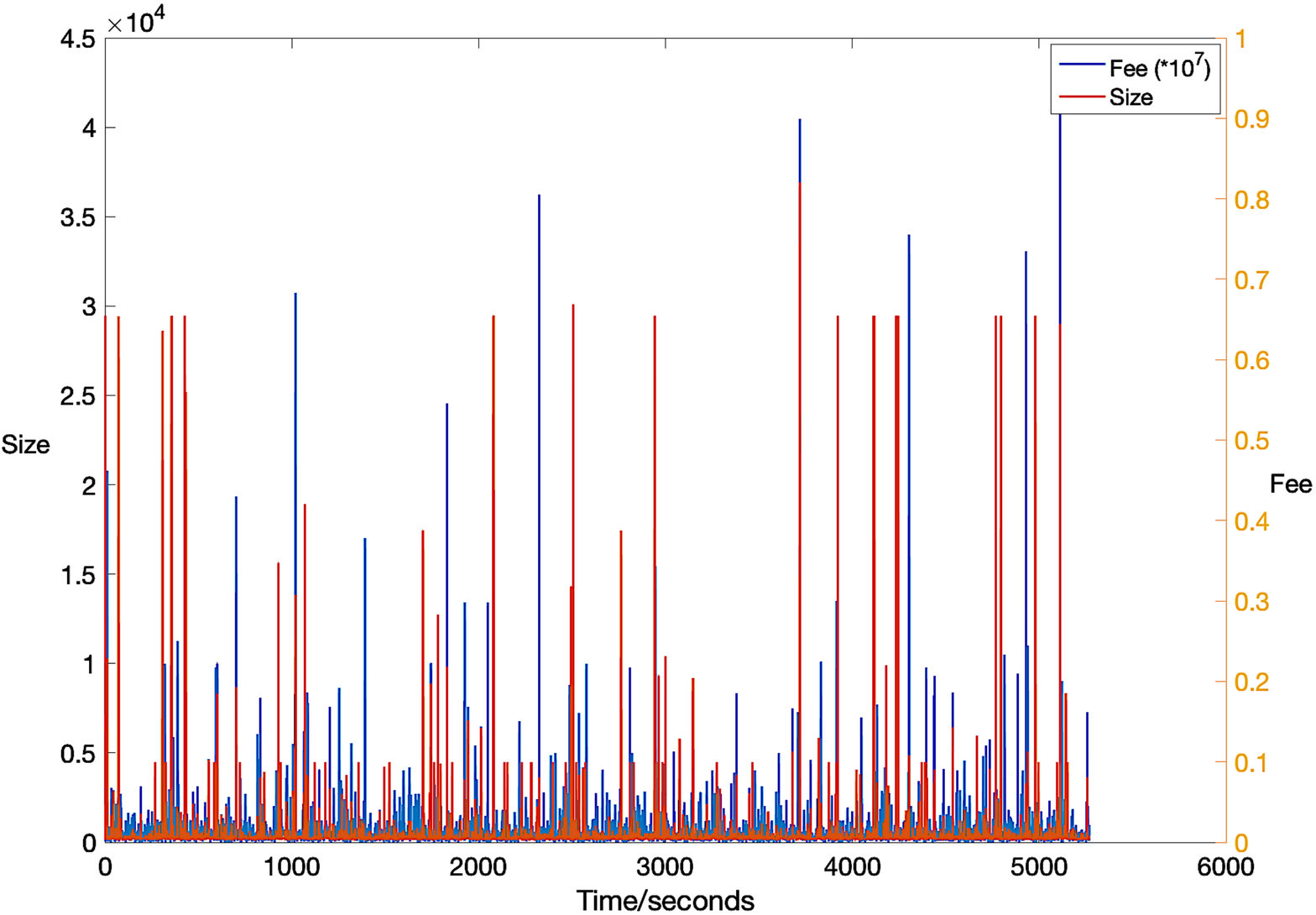}
  \caption{An overview of the transaction arrival process} 
  
  \label{sizeVsTraCom}
\end{figure}

In the remainder of this section, we focus on the transaction arrival process itself, which is characterized by transactions' inter-arrival times, and the size and fee of each transaction. Fig.\ref{sizeVsTraCom} provides an overview of this process, where 5000 unique transaction arrivals are ordered based on their arrival times $a_i$ recorded at our full node. 

\subsection{Transactions' Inter-Arrival Time}

\begin{figure}[ht!]
\centering
  \includegraphics[width=1\linewidth, height=0.4\linewidth]{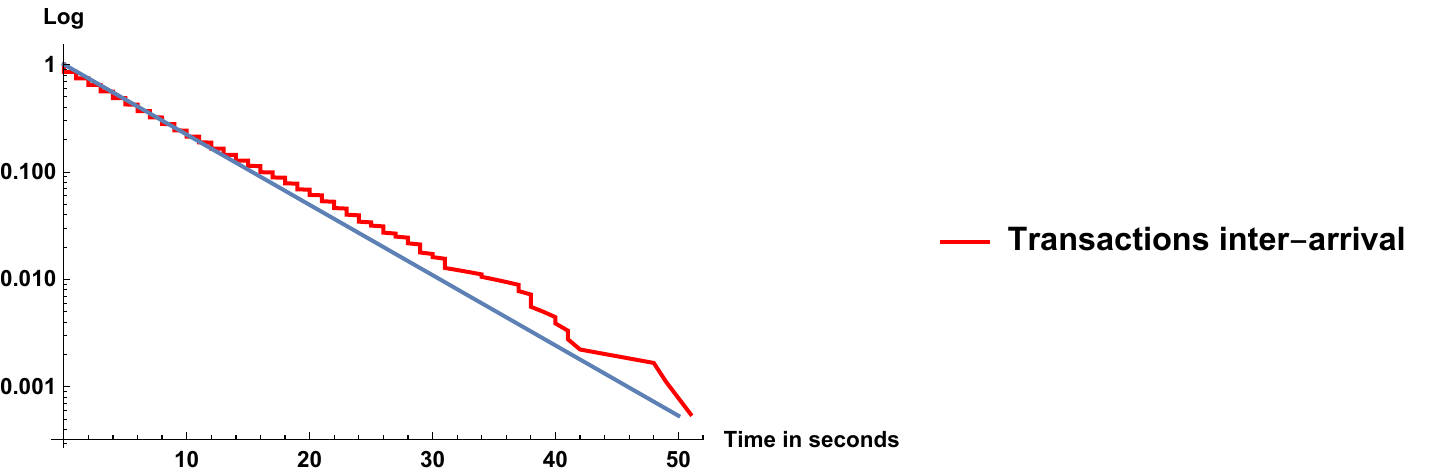}
  
  \caption{Distribution of transaction inter-arrival times, fitted with n.e.d}
  \label{transArrivallog}
\end{figure}

In the literature, it is often assumed that the transaction arrival process is a Poisson process. However, the validity of this assumption was not examined previously. To bridge this gap, a random period in the dataset was picked, which consists of 1861 transactions, and the inter-arrival time distribution of these transactions is illustrated in Fig. \ref{transArrivallog}.

As we can see from Fig. \ref{transArrivallog}, the transactions' inter-arrival times can be approximately fitted with an exponential distribution, which partially supports the Poisson arrival assumption. However, the figure also shows noticeable deviation. While the deviation for the CCDF value below $1\%$ may be attributed to the number of samples in this fitting test, the derivation is also visible for CCDF above $1\%$, which can hardly be found in the inter-block generation time and inter-block arrival time curves in Fig. \ref{block_gen} and Fig. \ref{block_arrival}.

\begin{finding}\label{f-4}
{\bf The transaction inter-arrival time may be approximated by an exponential distribution, but with noticeable deviation. }
\end{finding}

\subsection{Transaction Size and Fee}

According to the design of Bitcoin~\cite{Nakamoto}, how a miner selects transactions to form a block depends on the sizes $s_i$ and fees $f_i$ of transactions in the mempool. 
In Fig.~\ref{sizeVsTraCom}, an overview of them with regard to each transaction has been shown. To have a better understanding of them, we investigate their distributions and the correlation between them.

\subsubsection{Distributions of $s_i$ and $f_i$} Similar to distribution fitting for block size and fee introduced in Sec. \ref{fit}, it has also been conducted for transaction size $s_i$ and transaction fee $f_i$. Table~\ref{closefit_arrival} summarizes the distribution fitting results, where only the top three best fit distributions are included and  $\mu$ and $\sigma$ respectively denote the mean and standard deviation of the corresponding distribution. 

Different from block size and fee, Log-logistic, Log-normal, and Weibull are now the top three distributions.  As we can see from Table~\ref{closefit_arrival}, both $s_i$ and $f_i$ have similar close fit distributions, with log-logistic being the best for both. In addition, comparing the K-S values in Table~\ref{closefit} and Table~\ref{closefit_arrival}, it can be observed that even the best fit Log-logistic distributions for $s_i$ and $f_i$ still have non-negligible K-S value. This implies that further investigation is need to understand and find better ways to characterize the distributions of transaction $s_i$ and fee $f_i$.

A remark is that, while in Bitcoin, the default average transaction size is initially assumed to be 250 bytes, based on our analysis, the average size is almost twice of the default value. In addition, the size difference between small and large transactions is vast: while a small transaction only has 100 bytes, many transactions have a size of over 30 kilobytes.  

\begin{table}[ht!]
\caption{Distribution fitting of transaction size $s_i$ and fee $f_i$} 
\centering 
\begin{adjustbox}{width=0.48\textwidth}

  \begin{tabular}{|l|l|l|l|l|l|l|l|l|}
    \hline
    \multirow{2}{*}{Distribution} &
      \multicolumn{3}{c|}{Transaction size ($s_i$)} &
      \multicolumn{3}{c|}{Transaction fee ($f_i$)} \\
     & $KS$ & $AD$ & $Parameter$ & $KS$ & $AD$ & $Parameter$ \\
    \hline
    Log-logistics  & 0.2 &  225.1 & $\mu$= 5.42, $\sigma$= 0.31  & 0.099 &  74.9 & $\mu$= 10.1, $\sigma$= 0.53  \\
    \hline
    Log-normal  & 0.27 &  457.9 & $\mu$= 5.55, $\sigma$= 0.75  & 0.14 & 148.3 &$\mu$= 10.16, $\sigma$= 1.07 \\
    \hline
    Weibull  & 0.33 &  888.95 & $\mu$= 5.86, $\sigma$= 0.76  & 0.17 & 290.9 & $\mu$= 10.66, $\sigma$= 0.84 \\
    \hline
   
  \end{tabular}
  \end{adjustbox}
 \label{closefit_arrival}
\end{table}

\subsubsection{Joint distribution of $s_i$ and $f_i$} 
\begin{figure}[ht!]
\centering
  \includegraphics[width=1\linewidth, height=0.5\linewidth]{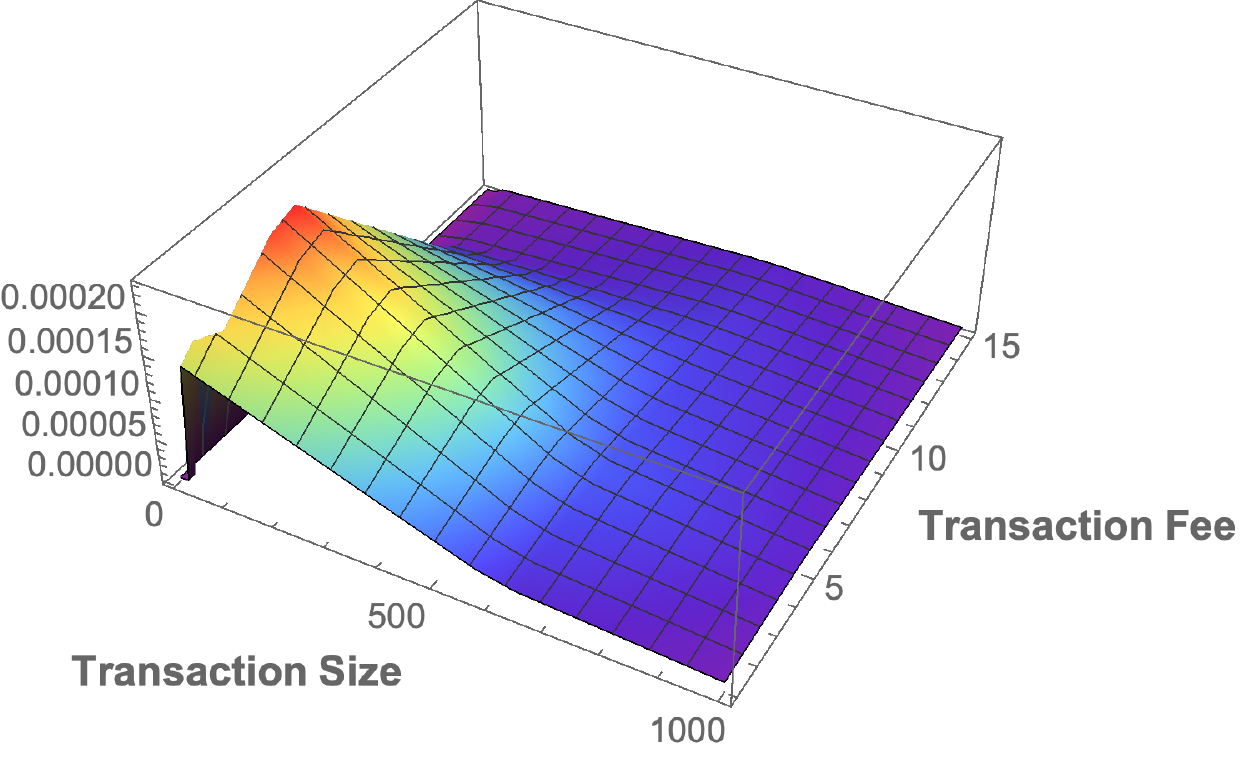}
  \caption{The joint PDF of $s_i$ and $f_i$}
  \label{JointcdfArrival}
\end{figure}

Fig.~\ref{sizeVsTraCom} shows that transaction size and fee do not seem to exhibit a clearly visible, strong / positive correlation. While some of the low fee transactions have high sizes $s_i$, we can also see transactions with higher fees having smaller transaction sizes.  To gain a more complete view, the joint distribution of $s_i$ and $f_i$ is investigated. For the same transactions shown in Fig.~\ref{sizeVsTraCom}, the joint distribution result is shown in Fig.~\ref{JointcdfArrival}. 

Fig.~\ref{JointcdfArrival} indicates that 90\% of the transactions have a size of not more than 500 bytes. But, in 1\% of the cases, the transaction size can be more than 30 kilobytes.  Similarly, the fee associated with each transaction is below 0.0006 BTC 90\% of the time, but it can grow higher than 0.001 BTC in 1\% of the cases.  The distribution shows that while there are a lot of small transactions, there is a significant fraction of tens and hundreds of transactions with a higher fee. Fig.~\ref{JointcdfArrival} also confirms that the correlation between transaction size and fee is weak.

\begin{finding}\label{f-6}
{\bf The correlation between the size of a transaction and its fee is unclear. }
\end{finding}

\section{Mining Pool Dynamics}  \label{sec-mem} 

In this section, the dynamics of the mining pool, which is affected by the transaction arrival process and underlays the block generation process, are investigated.

In Bitcoin, the transaction arrivals are collected and stored in the mempool to be picked up and included in a block. The mempool quantity can be adjusted based on the miner or full node runner's interest. Depending on the allocated mempool space, the full node running out of the allocated mempool capacity will block new transaction arrivals. The miners usually assign ample disk space to the mempool so as to have more choices in selecting transactions.  Our installed node has 35 MB size capacity; beyond this limit, arrivals will be discarded.  

\begin{figure}[t!]
\centering
  \includegraphics[width=1\linewidth, height=0.6\linewidth]{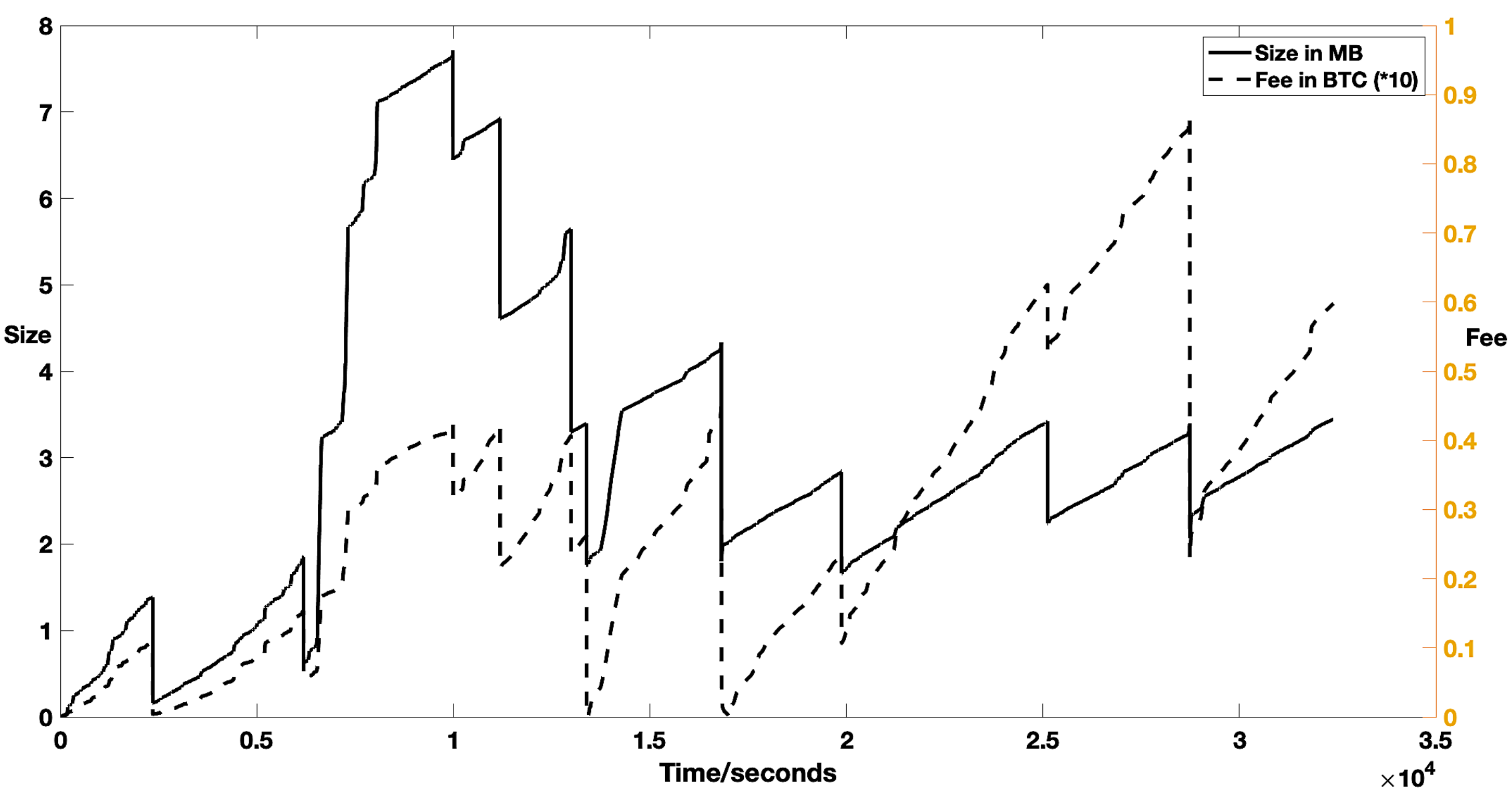}
  \caption{The mempool state change at block generation} 
  \label{mempool_content}
\end{figure}

A trace of the mempool size in terms of bytes and accumulated fee over ten-block formations is shown in Fig.~\ref{mempool_content}. The x-axis represents arrival times of the blocks, and the y-axis the accumulated entry size and fee, where the fee is multiplied by ten for better visibility. Each vertical descent in the size curve represents a new block formation and the hight of the descent implies the total size of transactions included in the block, i.e. the size of the block. The corresponding vertical descent in the fee curve represents the fee of the block.  

As indicated by Fig.~\ref{mempool_content}, the relationship between block size and block fee is not linear: a bigger  block  does not guarantee a higher fee fee and vice versa. When adding transactions into a block, higher priority may be given to the fee than to the number of transactions waiting for confirmation.  
For instance, we have observed there were often 5000 - 15000 transactions waiting,  while the blocks consider fee rather than the mempool size.  It is also visible that the mempool state has a fee close to zero at two times, implying that most transactions by then have been confirmed. However, we have also observed that these are low fee transactions that have to wait even longer time to be processed. If a transaction has a bigger size and small fee combination, it may occupy the memory space for a longer time before confirmation. 

\begin{figure}[t!]
\centering
  \includegraphics[width=1\linewidth, height=0.5\linewidth]{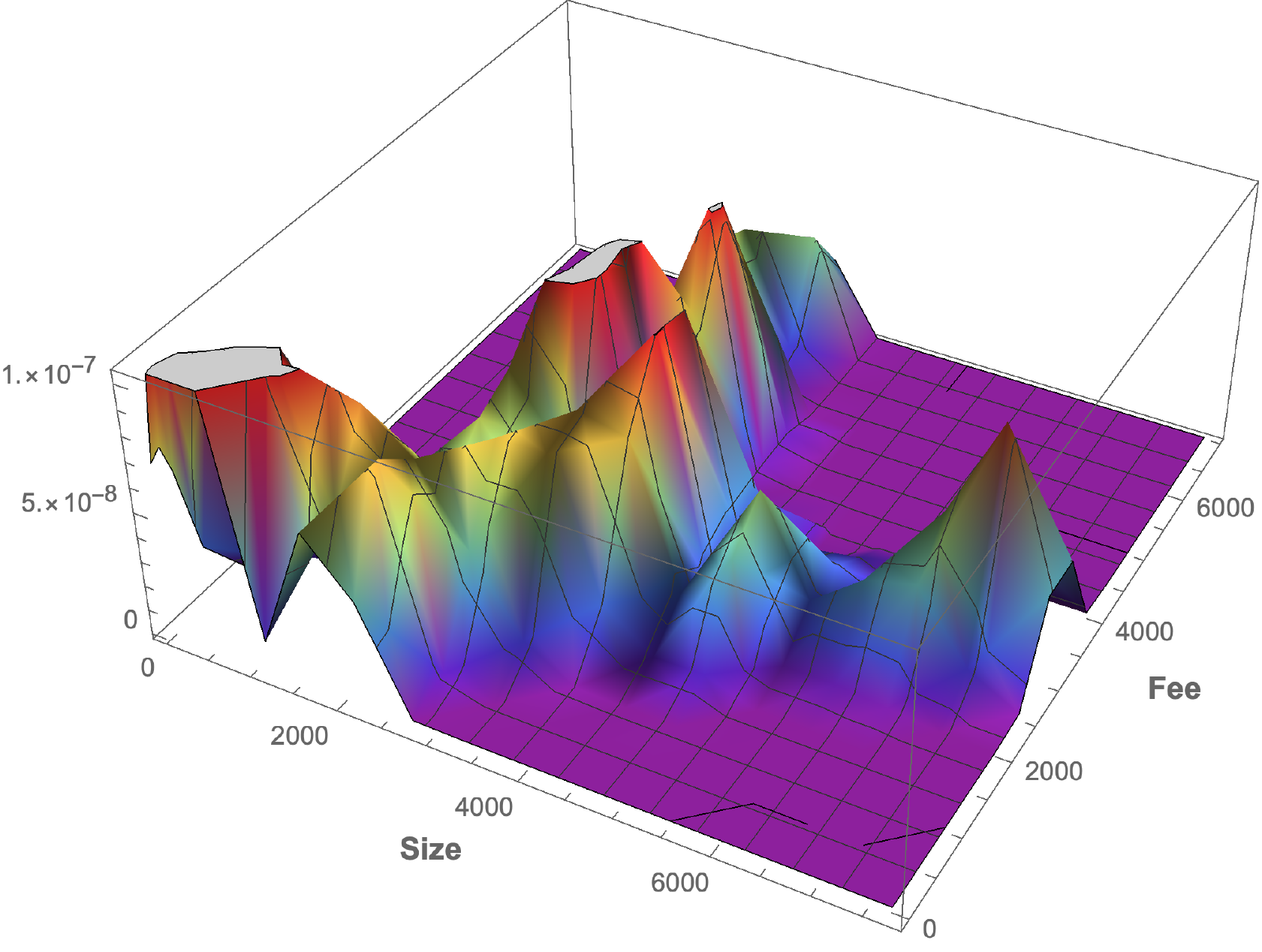}
  \caption{The joint PDF of instantaneous fee and size in the mempool}
  \label{jointmempool}
\end{figure}

Fig.~\ref{jointmempool} shows the joint PDF of accumulated fee and accumulated transaction size in the mempool.  The figure demonstrates that there is a strong correlation between them.  
Note that, the miner's financial interest encourages it to choose transactions with a high fee per byte associated with them so as to maximize the benefit, i.e. the fee, of the block. This gives an explanation about the strong correlation. 

For the same reason, in the literature, a fee-based priority queueing model has been simply assumed for the mempool [4] [5]. 
However, this assumption is too coarse to explain what are shown in Fig.~\ref{mempool_content} and Fig.~\ref{jointmempool}. 
For instance, Fig.~\ref{mempool_content}  shows some blocks contain only a few transactions while the block size is filled to the maximum, implying that in these cases, transaction size seems to have been prioritized rather than fee.

\begin{finding}\label{f-7}
{\bf A simple fee-based priority queueing model  cannot well capture the dynamics of the mempool. }
\end{finding}

\section{Related work}\label{related}
Most of the available Bitcoin literature focuses on analyzing transaction identity, including~\cite{denoBitcoin}~\cite{bitp2p}~\cite{FistfulBitcoin}~\cite{EOSIOBlockchain}. Other than the various Bitcoin platforms e.g. those included in Table \ref{sourcecomparison}, only a few research works have investigated the transaction data residing on Bitcoin's longest chain.  Pavithran et al.~\cite{bitcoinAnalyse} collected data ranging from 2009 to 2013 to study the systems long-term credibility.  
Wu et al.~\cite{ForensicAnalysisBitcoin} developed a model that translates the Bitcoin transaction data into a formal model that focuses on finding bitcoin addresses involved in specific transaction patterns. Other than the transaction source pattern, to know the client information,  Guo et al.~\cite{ TowardsTracingBitcoin}~\cite{LN} prepared a small testbed containing a server and three client instances.  The testbed was used as connectivity that enables the network traffic collector, in this case, the Wireshark, to collect the traffics between the nodes. 

Gervais er al.~\cite{bitcoinPerf} developed a framework that enables to conduct of analysis on the impact of block size and inter-arrival time over the ledgers security state. 
Even with the possibility of increasing the block size, it was demonstrated in~\cite{trasactionConfirmation}, that even with a high fee, transactions exhibit high confirmation times. 
Göbel et al.~\cite{BitcoinBlockchainDynamics} also illustrated the same concept by assuming that miners perform hash calculations independent of the previous hash calculations, implying geometric distribution for the success of the trial which can be approximated by an exponential distribution.  Similarly, in~\cite{batchqueuing}, a batch processing queueing system is developed which uses numerical and trace-driven simulation to validate the assumptions and results.

In brief, the state of the art shows a clear gap in characterizing the transaction and block processes, and in addressing the relationship between block size, miner's incentive, the number of transactions per block, the mempool status, the transaction inter-arrival time, and their impact on each another. In particular, to the best of our knowledge, no previous work has combined both global block information on the blockchain and local transaction / block information that is available only at a Bitcoin node in studying the transaction and block characteristics. Our work has been intended to fill this gap.

\section{Conclusion} \label{sec-con}
Through analyzing the data collected from a measurement setup, which contains transaction and blcock information both on the blockchain and from the node, we presented a comprehensive study on the transaction characteristics of Bitcoin. A set of new results and findings have been reported, including examining the validity of several hypotheses /  assumptions used in the literature. 

Specifically, for exponentially distributed inter-block  generation / arrival times, we found that the two literature hypotheses cannot be justified by the measurement, and it is likely attributed to exponentially distributed block generation at major miners. In addition, for transaction inter-arrival time, though its distribution may be approximated with an exponential distribution, there is  noticeable deviation. Besides, for characterizing the mining pool, no convincing evidence has been found to support the fee-based priority queueing model. Furthermore, while the size of a block and the number of transactions in it exhibit clear strong correlation, transaction size and fee seem to be more independent.

 
As a highlight, the idea of involving the mempool in the measurement, in addition to the commonly used ledger information, has enabled us to study the transaction characteristics of Bitcoin and find the fundamental relationships among the core features. As a future work, we will investigate how to exploit this idea to manage the mempool to improve the throughput and reduce transaction waiting time while keeping the current block size limit.


\AtNextBibliography{\small}
{\tiny \printbibliography}

\end{document}